\documentclass[11pt,draftcls,onecolumn]{IEEEtran}

\usepackage{graphicx}
\usepackage{tabulary}
\usepackage{multirow}
\usepackage[center]{caption}
\usepackage[hyphens]{url}
\usepackage{array}

\ifCLASSOPTIONcompsoc
\else
\fi

\usepackage{epstopdf}
\usepackage[cmex10,fleqn]{amsmath}

\DeclareGraphicsExtensions{.eps}

\begin{document}

\title{Could We Distinguish Child Users from Adults Using Keystroke Dynamics? }

\author{Yasin~Uzun,
		Kemal~Bicakci,
		Yusuf~Uzunay~
\IEEEcompsocitemizethanks{\IEEEcompsocthanksitem Y. Uzun and Y. Uzunay are with Middle East Technical University, Ankara, Turkey
E-mail: yuzun@epdk.gov.tr, yuzunay@gmail.com.
\IEEEcompsocthanksitem K. Bicakci is with TOBB University of Economics and Technology, Ankara, Turkey
E-mail: bicakci@etu.edu.tr.}
\thanks{}}

%

\IEEEcompsoctitleabstractindextext{%
\begin{abstract}
Significant portion of contemporary computer users are children, who are vulnerable to threats coming from the Internet. To protect children from such threats, in this study, we investigate how successfully typing data can be used to distinguish children from adults. For this purpose, we collect a dataset comprising keystroke data of 100 users and show that distinguishing child Internet users from adults is possible using Keystroke Dynamics with equal error rates less than 10 percent. However the error rates increase significantly when there are impostors in the system.

\end{abstract}

}

\maketitle

\IEEEdisplaynotcompsoctitleabstractindextext

%
\IEEEpeerreviewmaketitle

\section*{Acknowledgment}

We thank all participants who participated in the experiments. We would like to thank TUBITAK (The Scientific and Technological Research Council of Turkey) for providing financial support to Yasin UZUN during his PhD study. We thank anonymous reviewers for their helpful suggestions.

\section{Introduction}
%
%

%
%
%
%
\IEEEPARstart{C}{omputer} usage age is decreasing rapidly. An evidence for this trend is the popularity of social networks among youngsters. Majority of young individuals use computers for communicating with their peers on social networks. Although Internet has the benefits, as we all know, it is also a source of many threats especially for children.

In parallel to widespread penetration of Internet, a serious debate has emerged on the online content that may negatively affect the morality of children. In a survey conducted in India, it is revealed that 67\% of the children under 10 had a Facebook account before they were 10 and 82\% of them received inappropriate messages \cite{india}. Incidents like these cause families to approach Internet with severe suspect, but many of them do not know how to react appropriately. As a response, governmental authorities are actively trying to protect youngsters from the possible threats of Internet.

Although there is a considerable effort for protecting children from harmful content and Internet threats, most of them are based on shutting down certain domains. As a result, the controls are either too restrictive that they also distract adult users or they do not provide sufficient level of production. They also have privacy related problems. We believe that children will be better protected without distracting adult individuals if there is a way to differentiate children and adults on computers automatically.

An example application for age group detection is a children-only social network site where adults are not allowed to access. With such a functionality, perpetrators and criminals cannot get involved with minors using these networks. Another potential application area may be police investigation cases for identifying criminals, who introduce themselves as youngsters on online chat applications to abuse minors. While a policeman is chatting with an individual on the other side, if he is suspicious that the person is a potential criminal, who is imitating a child, he may use such an application to get a hint about the age of the person.

Other than forensic applications, age group detection applications may be used for commercial purposes. Suppose that a potential customer browses a web site that makes online sales. If it is possible to make a prediction about the age group of the customer, the web site can make product recommendations based on his/her age group. But significant care should be paid to privacy issues for such an application, to avoid possible legal disputes.

As a summary, we can list the benefits of automatically identifying the age group (minor or adult) of online Internet users as follows:
\begin{itemize}
\item For web domains which may be harmful to children, an access control mechanism can be built, which becomes active only for child users.
\item Private domains for exclusive use of children can be built, where adults are not allowed to access.
\item Software tools can be implemented for criminal investigations, which detect the perpetrators who falsely introduce themselves as minors.
\item It may be possible to display content suited for the age group of users, in commercial web sites.
\end{itemize}

We conjecture that distinguishing child computer users from adults is possible by analyzing typing behavior of users. It is already demonstrated that Keystroke Dynamics can be successfully used for identity verification. In this study, we are interested in a novel application area of Keystroke Dynamics; detecting the age group of users. Our problem can be formally defined as follows: Given a training set of typing patterns consisting of interkey latencies, where each pattern is assigned to one of two labels: adult and child. The goal is to find the relation that maps input patterns to one of the given labels. More specifically, our problem is to find a relation between keystroke data and the age group of typists where there are two age groups; children are defined as the users under age 15 and adults are the individuals above age 17. To get discrete set of users, teenagers (age 15-17), who are hard to fit either of these two groups, are excluded in this study. The typing data consists of numerical elements, which correspond to time periods in microseconds that elapse between consecutive key press events and time between key press and key release events.

To solve the problem, we first collect a dataset comprising keystroke data of 100 users because available public datasets do not contain the age information. To enable future studies, we make the dataset publicly available together with our implementation \cite{dataset}. We show that distinguishing child Internet users from adults is possible using Keystroke Dynamics with an equal error rate of 8.8 percent. On the other hand, we also show that Keystroke Dynamics based age group detection is vulnerable to non-zero effort attacks.


\section{RELATED WORK}



Keystroke Dynamics is the process of monitoring and analyzing the typing behavior of users on a computer keyboard and is extensively studied in last two decades, with the main focus on user authentication. Different methodologies are proposed in these studies, tested on the collected datasets and error rates are reported with respect to different metrics.

There were many studies on Keystrokes Dynamics, but usually it was not possible to compare their success  rates since the datasets collected in the studies were different and were not open to public. Realizing this deficiency, Killourhy and Maxion  \cite{anomaly} collected a keystroke benchmark dataset from 51 subjects, who typed the same password for 400 times. They implemented 14 different algorithms previously used in the literature and compared the equal error rates on this dataset. They opened their dataset to public together with source code of algorithms implemented in R scripting language  \cite{rlang}. This has provided a basis for comparative future studies on the same dataset.

A follow-up study that is made on the benchmark dataset of Killhourry and Maxion \cite{anomaly} is for analyzing the success of neural networks for Keystroke Dynamics \cite{second}. In this study, neural networks were run with various learning algorithms. In addition, negative examples, which refer to the samples that belong to subjects other than the genuine user, are also used in training phase. The resulting error rates show that neural networks can outperform other algorithms, when run with suitable configuration.

The security (attack-resistance) of Keystroke Dynamics as a way of user authentication was also studied. Assuming that keystroke latencies are compromised by the attacker, the authors emulated attacks on keystroke based verifiers, and reported that 87.75 percent of the forgeries were successful \cite{snoop}. It has been argued recently that algorithmic forgeries based on snooped keystrokes drastically increase error rates of keystroke verification systems \cite{rahman2013snoop}.  In another study \cite{traits}, keystroke based security systems were shown to be vulnerable to synthetic impostor attacks based on general typing habits.


Although studies in  Keystroke Dynamics literature have the main focus of verification and identification, it is also pointed that using typing data for extracting demographic information could be an interesting application \cite{application}. However, to our knowledge, the only study that focused on such an application is the work of Giot and Rosenberger  \cite{giot}, in which they used typing data to predict the gender information of individuals. The authors used support vector machine to classify male and female typing patterns and reported a success rate of 91\%.


To our knowledge, there was no earlier study on automatically predicting the age group of an individual from his/her behavioral biometrics. But there has been considerable research effort on extracting information about personal characteristics from handwriting. In fact, these studies have matured and formed the discipline of Graphology  \cite{graphology}, in which human or machine interpreters evaluate the handwriting of individuals to extract demographic information.




There is a recent study \cite{fingerprint} on age estimation through fingerprint, a commonly used physiological biometrics. In this study, the authors used 3570 fingerprint images, which were divided into 5 age groups. Using Discrete Wavelet Transform and Singular Value Decomposition, feature vectors are extracted from fingerprints. Then, k-nearest neighbor algorithm is used for classifying feature vectors. The authors reported 76.84\% success rate for male and 59.26\% for female in this study.

We believe that an application that classifies computer users according to their age groups using typing data is an interesting new research challenge. The number and diversity of studies already performed on Keystroke Dynamics encourage us for working on distinguishing age groups based on keystroke information.

\section{DATASET}
To our knowledge, there is no available dataset to test the feasibility for inferring the age group of computer users from their keyboard use because public datasets do not contain age information. Therefore we collected our own dataset for this purpose. Our dataset consists of typing samples collected from 100 subjects. Subjects are equally distributed among four groups (25 for each): adult male, adult female, child male and child female. Our experimental work was performed with the approval of Middle East Technical University, Human Subjects Ethics Committee. For children subjects, written consent were taken from their parents.

To enable future studies, we make the dataset of the adults publicly available \cite{dataset}, in deidentified format for privacy reasons.

\begin{figure}[!htb]
\centering
\includegraphics[width=3.5in]{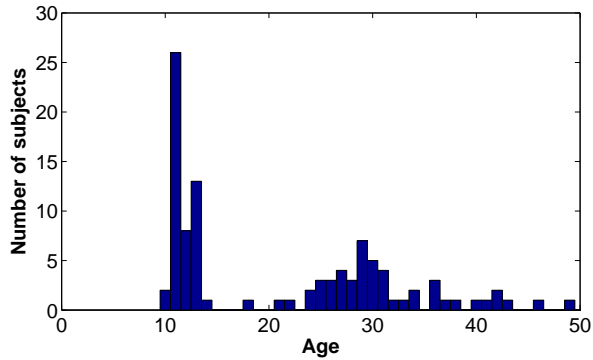}
\caption{The histogram showing the distribution of subject population with respect to age.}
\label{fgr:histogram}
\end{figure}

All the recruited subjects were free from orthopedic problems, which can possibly cause disturbance during typing. All of the subjects have basic computer literacy skills, such as using mouse, keyboard and an X-Windows application. The histogram plot for ages of subjects is given in Figure~\ref{fgr:histogram}. Ages of the adults vary between 18 and 49 while ages of the children vary between 10 and 14.

The data collection is performed using the same laptop computer, which is HP Compaq 6000 Pro SFF PC with Intel Core i5 CPU M430 @ 2.27 GHz processor having Microsoft Windows XP Professional SP2 operating system.  All users were provided the same external Turkish QWERTY keyboard (A4 Tech Kr-73), which is similar to an English QWERTY keyboard, but also contains 6 additional Turkish characters that do not exist in the English alphabet on the right part of the keypad. It is shown that changing keyboard does not significantly affect accuracy \cite{keyboard} for Keystroke Dynamics based user verification\footnote{We leave the problem of analyzing the impact of keyboard change on the performance of keystroke based age group detection as a future work.}. During the experiments all the subjects were provided a comfortable chair and table in noise-free environment, to minimize the external effects that could disturb the typical typing behavior.

To collect data from the subjects, we developed a Windows application developed in Visual Studio.NET 2008 programming environment, which we also make publicly available \cite{dataset}. The application consists of series of forms appearing in the predefined order. All the raw data collected during the experiments is recorded in a Microsoft Office Access 2007 database.

All the subjects were informed about the experimental procedure in short but the purpose of the experiment is not mentioned in order to avoid any effection on their natural typing behavior. Initially, the subjects were greeted by a welcome screen, in which they entered their name, surname, gender, year of birth and class (for primary school students only)  information. The only information that is required for our current work is the year of age, but we also retained name and surname in order to recognize a subject in case they may provide data at future sessions and also for future studies. In case the subjects may be uncomfortable about revealing their identities, they were informed that they can use nicknames in this screen, but none of them preferred to do that. Lastly, gender data is used to build a dataset with equal number of males and females in each age group.

In the next step, the users were confronted with a short survey including questions that may be related to their typing behavior. First question was whether the user is left or right handed. In parallel to the general human population, majority of the participants are right handed in both age groups (5 users among children and 4 users among adults are left-handed).



The second question was whether the participants own a personal computer. This question is prepared to learn the familiarity of subjects with computers. Only 13 of child subjects and 3 of the adult subjects declare that they do not have a computer.



Next, we questioned the experience of the subjects on computer use. 19 children have been using computers for less than one year, while there is no such an adult participant. 21 children and a single adult have been using computers for more than 1 but less than 5 years. Remaining participants have been using computer for more than 5 years.



Besides experience, frequency of computer use may also affect the typing style of individuals. Hence, we asked the subjects their daily computer usage time on average. Among children, 34 of them spend less than one hour on the computer per day while this is true for only 2 adults. 13 participants from the children group and 4 participants from the adult group spend 1 to 4 hours on the computer each day. Remaining subjects use computers more than 4 hours per day.



Typing proficiency is not always directly related to the time spent in front of computers, since some times users rarely use the keyboard. Hence, the subjects were also asked to predict the number of words they type in a day on average. 16 of the child participants and 4 of the adult participants declared that they type less than 20 words per day. 20 of the child participants and 16 of the adult participants type between 20 and 200, while the remaining type more than 200 words.



Once the survey is completed, the users are directed to the typing screen. In the experiments, we used fixed text approach in which the participants copied the text appearing on the screen into a textbox. Most of the real life online activity happens with free text, however,  the effects of typing differences between free and fixed text is shown to be insignificant \cite{free}. Therefore, fixed text is used for collecting the dataset instead of free text, since it takes much more time and effort to perform the experiments with free text.


First, the subjects are requested to type the Turkish phrase; ``Mercan Otu'' which means ``Coral Grass'' in English. We label this dataset as ``Turkish dataset''. This dataset corresponds to relatively easy typing task. We also note that our choice of the phrase ``Mercan Otu'' has the following reasons. Firstly, the phrase should be long enough to produce a feature vector of keystroke events. So it is decided to choose a password of 10 characters, in parallel to work of Killhourry and Maxion \cite{anomaly}. Second, our desire was that the subjects type the phrase without error, otherwise if they delete some characters it would not be possible to generate a fixed length feature vector due to varying number of keystrokes.

The phrase ``Mercan Otu'', has some additional nice properties. Although the phrase is Turkish, it does not contain any characters not present in English alphabet. We think this may provide the opportunity to perform comparative studies in different countries in which the same phrase may be typed by the subjects. We think interesting results may be obtained in such studies. Another desirable property of the chosen phrase is that it consists of two words, therefore users have to hit the space key, which may be important  for capturing typing behavior. Furthermore, the initial characters of words are in capital, forcing users to show their typing habit for capital letters.

Subjects were requested to type the Turkish phrase and finish each typing session by hitting the carriage return (Enter key). The phrase was visible in a textbox on the screen. The subjects were warned not to use backspace and delete characters. For erronous entrance, we have a ``RESTART'' button on the form, which clears the textbox and restarts the session. Despite the warning, users accidentally hit backspace and delete keys. In those cases, they were confronted with a warning message reminding not to use these keys, the textbox is cleared and the session is restarted. Another problem could be incorrect inputs subjects were not aware of. The application ensures that the given input matches the desired phrase case-sensitively; otherwise the last typing session is discarded. If the phrase is typed correctly without using deletion keys, the input is accepted, the timestamps of the key events are recorded in the database and session counter (which is displayed to the subject during typing) is incremented. The subjects were requested to type the text for 5 times.

While the subjects type on the keyboard, the software application keeps the key press and key release event timestamps in the memory. GetTimestamp() procedure of the Stopwatch Class of Microsoft .NET Framework is used to capture the timestamps of the key events with the accuracy of microseconds. We also force the application to use the second core of the processor and assign highest priority to the running thread in the implementation, in order to prevent operating system interruptions.

After completing the desired number of successful typing entries, the subjects are greeted with a message telling that they finished the first step successfully. In the next step, they were requested to type a password ( ``.tie5Roanl'' ) of ten characters, which is used in a previous study  \cite{anomaly}. We label this dataset as ``Password dataset'' and believe that this same choice enables further comparative studies on Keystroke Dynamics. The subjects typed this phrase in the same way as the first one, with the same number of times. At the end, subjects were offered a bar of chocolate to appreciate their efforts. Some of the adult subjects asked about the purpose of the experiment after the process and developed significant enthusiasm about the results of the study after learning about our research objectives.

The collected dataset consists of 1000 typing samples belonging to 100 subjects. For each subject, there are 10 typing samples, of which 5 are for Turkish phrase and remaining 5 are for password phrase. Each sample is represented as a feature vector with 5 header elements and 31 data elements. The header fields are user id, gender, age group, year of birth and session number, which are all represented with integers. The data elements correspond to time periods between key events in microseconds in the dataset.

There are several alternatives for selecting the feature set. Most commonly used metrics are digraph, which is the amount of time elapsed between the key events for two consecutive characters and trigraph which is the same measurement for three consecutive characters. In this study, digraph measure is used to compose the dataset, as preferred in majority of the studies for Keystroke Dynamics. In order to make comparative studies with the Keystroke Dynamics Benchmark Dataset \cite{anomaly}, the dataset is composed in an identical format. In each feature vector, 11 of the data elements are key duration times (the amount of time the key is pressed for each character including Enter key), 10 of them represent the amount of time between key press events between consecutive characters and remaining 10 are the time values between key release and key press events for consecutive characters.

The feature vector for the data elements is depicted in Equation \ref{eq:feature}, in which key press and release events for the character `M' are denoted as \emph{P(M)} and \emph{R(M)}, respectively. In the equation, \emph{P(M)P(e)} corresponds to the time period between the key press event when typing letter `M' and the key press event when typing the letter `e'. In the same manner, \emph{R(M)P(e)} corresponds to the time period between the key release event when typing letter `M' and the key press event when typing the letter `e'. The time period between the key press and release events for the same key `M', which is the duration time for that key, is shortened as \emph{D(M)}.

\begin{eqnarray}
\label{eq:feature}
&& D(M), \mbox{  } P(M)P(e), \mbox{  }  R(M)P(e), ...  , D(u), \mbox{  }  P(u) P(Enter), \mbox{  }  R(u)P(Enter), D(Enter)
\end{eqnarray}


\section{ANALYSIS}

Using the collected dataset, we attempt to analyze the performance of discriminating child typing samples from adult typing samples. For this purpose, we employ common distance metrics and pattern recognition techniques that are frequently used in Keystroke Dynamics studies. We also employ artificial neural networks with different learning algorithms, which were previously tested for verification \cite{second}. In our dataset, there are two groups of samples as children and adults where each group is divided into training and test sets. The training set is used to capture the domain knowledge while the test set is used to assess the classification accuracy for the proposed methodologies.

Selection of training samples may affect the success rates since some samples may better represent the domain while others do not. In order to neutralize this uncontrolled factor, we use 5-folds cross validation technique while computing the error. The dataset is initially divided into 5 equal subsets. In each iteration, one of the subsets is selected as the test set and the remaining 4 subsets are used for training. The process is repeated for 5 times, with an alternating test set. At the end, the average of the 5 test runs is computed to find the final value. The important point we keep in mind during subset division is to keep the samples of an individual in the same subset. Otherwise, samples from a single individual may be present in both training and test subsets, and it will not be possible to understand whether the method learns the subject or age group behavior.

We run our test scripts in  MATLAB numerical computing environment \cite{matlab}. To provide transparency of our test procedure and to promote future studies that may benefit from our study, we make all the source code (data collection application and MATLAB functions and scripts) of our implementation publicly available \cite{dataset}. Our test implementation is generic and can be used for testing any biometric dataset for binary classification, as long as the data is represented as composition of fix-length feature vectors \cite{dataset}.



The first metric we implement is the simplest one; the total time that elapses during the whole typing period. The general trend about typing speed is that children type slower than adults. Therefore an intuitive guess about the age of a typist could depend on speed of the typist. In order to apply this measure, we compute the total typing time for each session in training data (in microseconds), compute the mean vectors for adult and child group and calculate the sum of the elements of these two mean vectors separately. In test phase, the sum of the elements is computed and the absolute difference between this value and the sum of the adult and child mean vectors is separately calculated. If the difference is smaller for the adult mean, the sample is assigned as an adult; otherwise, it is assigned as a child.

\begin{figure}[!htb]
\centering
\includegraphics[width=3in]{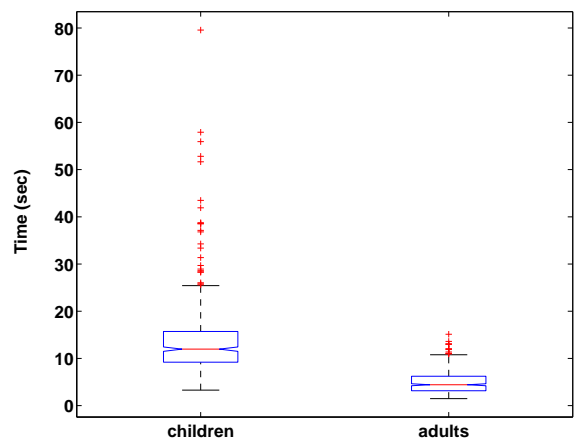}
\caption{Boxplot diagram showing the distribution of total typing time for two phrases in child and adult age groups.}
\label{fgr:anova1}
\end{figure}

To get an idea about the typing speed of the test participants, in Figure ~\ref{fgr:anova1}, we depict the box diagram showing the results of analysis of variance for the total time required to complete both phrases (Turkish phrase and the password). The values are divided into four subsets of same size with three separating points: lower quartile, median and upper quartile. The median values are shown with the horizontal lines inside boxes. The lower edge of the boxes (lower quartile) divide the values below the median value to two subsets of equal size and the upper edge (upper quartile) serves the same purpose for the values above the median. The whiskers, which are shown with dashed lines show the time ranges of 1.5 times the range between two quartiles from the ends of the boxes. The plus signs stand for the outlier elements which do not fit within boxes or whiskers. It is clear from the figure that the time values for the adults are condensed in a short range whereas a highly scattered distribution is observed for the children.

The second distance metric we use is squared Euclidean distance, which is the sum of the squared differences between the corresponding elements of two different vectors \cite{duda}. In this method, we compute the mean feature vector for the training set for both age groups as the first step. In test phase, squared Euclidean distance is measured between the incoming sample and both of the mean vectors of adult and child group. The sample is assigned to the group having the minimum distance.

The last distance measure we use is manhattan (city-block) distance metric \cite{duda}. At first step, mean feature vectors are computed in the same way as Euclidean distance. During testing, the manhattan distance value, which is the sum of the absolute differences between the corresponding elements of two vectors is computed for the new sample and each of the two (children and adult) mean vectors. The sample is labeled with the label of the group that has lower manhattan distance value to itself.


With the nearest neighbor classification, the only work in learning phase is to store the training sample points together with their class labels. When a new sample is to be classified in test phase, the algorithm searches the $k$ number of nearest neighbors, which is a user defined parameter (selected as $3$ in our work). The sample is assigned with the label of the group of training patterns with larger number of elements in the neighborhood.



Another classification method we use is linear discriminant analysis which constructs discriminant equations using the feature vector elements as input parameter by maximizing the difference between the classes \cite{duda}.






We use two SVM implementations. In the first, a linear support vector machine is trained using the set of training patterns, which are vectors composed of keystroke latencies. In the second, we use SVM with Gaussian radial basis function (RBF) kernel mapping \cite{tutorial}.





With neural networks, as the method for calculating weight values between neurons, we employ backpropagation algorithm in our study. There are different alternatives for using the gradient value for updating weight values. In this study, we perform tests using six of these methods (one from each family of algorithms): gradient descent with adaptive learning rate, conjugate gradient backpropagation with Fletcher-Reeves (FR) updates, BFGS quasi-Newton method, one-step secant backpropagation, scaled conjugate gradient backpropagations, Levenberg-Marquardt backpropagation \cite{second} \cite{toolbox}.


We design a neural network having three layers for our experiments: input layer, hidden layer and output layer. The number of neurons in the input layer is set to the feature vector size, the hidden layer size is two thirds of the size of input layer (a rule of thumb) and there is a single output neuron.  In learning phase, we first set the initial weights, randomly from a uniform distribution of real values (0, 1). Then, training samples are given as inputs to the neural network with the outputs set to +1 for adult typing samples and -1 for child typing samples. In the test step, feature vectors of the test samples are used as input signals for the network. If the result of the output neuron is greater than (or equal to) zero, the sample is labeled as an adult sample, otherwise as a child sample.

\section{Results}

We test and compare the classification methods described in the previous section for classifying typing samples of adult and children participants. There are two types of error. Type-1 error is the ratio of adult typing samples mislabeled as child sample to all adult typing samples and type-2 error is the ratio of child typing samples mislabeled as adult sample to all child typing samples. In our work, we report equal error rates, the value obtained when type I and type II error rates are equal.


Our dataset is divided into training and test sets. The algorithms capture the typing behaviors of age groups by using the training set and the success of the algorithm is assessed by computing the ratio of correctly labeled test samples to the test set size.

We measure the performance of the algorithms for three different datasets. Our first dataset consists of interkey times for Turkish phrase (``Mercan Otu''). Although the subjects were not familiar with this phrase before, the words ``Mercan'' and ``Otu'' are well known common words in Turkish. As a result, the subjects did not need to look at the screen during writing. Our second dataset contains the interkey time values for password phrase (``.tie5Roanl''). Typing this password phrase is a relatively  harder task for the subjects, for which they occasionally had to look at the screen.

Our last dataset is  concatenation of the two datasets. In this dataset the $n^{th}$ feature vector (sample row) is concatenation of the  $n^{th}$ feature vector of the Turkish dataset and the password  dataset. Since Turkish and password datasets are sorted by subject and session order, this last dataset represents what feature vector we would collect if the subjects typed two phrases consecutively. Hence it reflects the behavior for typing meaningful and password like phrases together. For all three datasets mentioned above, we compute equal error rates using receiver operating characteristic (ROC) curves. The test results are listed in Table \ref{testresults}.

\begin{table*}[!htb]
\caption{The performance of tested algorithms for discriminating age groups.}
\label{testresults}
\centering
\renewcommand{\arraystretch}{1.2}
\begin{tabular}{|p{5.5cm}|>{\centering\arraybackslash}m{1in}|>{\centering\arraybackslash}m{1in}|>{\centering\arraybackslash}m{1in}|}
\hline
  & \multicolumn{3}{c|}{\textbf{Equal Error Rate (\%) }}
\\ \hline
  \textbf{Algorithm} & \textbf{Turkish dataset}  & \textbf{Password dataset}   & \textbf{Concatenated dataset} \\ \hline
Speed (Total time)	&	10.4	&	16.4	&	12.4	\\	\hline
Euclidean distance	&	10.4	&	16.0	&	12.8	\\	\hline
Manhattan distance	&	10.4	&	16.4	&	12.0	\\	\hline
Nearest neighbor	&	10.8	&	16.0	&	12.0	\\	\hline
Linear discriminant analysis	&	10.0	&	\textbf{11.6}	&	9.6	\\	\hline
Support vector machine (Linear) 	&	\textbf{8.8}	&	12.8	&	10.0	\\	\hline
Support vector machine (RBF)	&	11.6	&	13.8	&	12.0	\\	\hline
Gradient descent bp. 	&	27.2	&	30.4	&	39.2	\\	\hline
Conjugate gr. bp. with FR updates 	&	10.8	&	12.8	&	9.6	\\	\hline
BFGS quasi-Newton bp.	&	10.4	&	13.6	&	\textbf{8.8}	\\	\hline
One step secant bp.	&	10.4	&	12.8	&	9.6	\\	\hline
Scaled conjugate gradient bp.	&	10.0	&	13.2	&	10.8	\\	\hline
Levenberg-Marquardt bp.	&	12.8	&	14.4	&	11.2	\\	\hline

\end{tabular}
\end{table*}

The equal error rates are always lower for Turkish dataset than those for password dataset, showing that classifiers are better at recognizing the behavior of the participants when they type meaningful phrases. A possible reason behind this result is that the participants had to look at the screen while typing the password phrase and had a smoother typing behavior. The equal error rates for concatenated dataset are always lower than those obtained for the password dataset, hence, it can be said that password dataset is the worst alternative for discriminating age groups. However, it is hard to decide the best dataset since error rates for four algorithms are lower with concatenated dataset while the error rates for the remaining algorithms are lower when Turkish dataset is chosen.

For Turkish dataset, minimum error rate is 8.8\% and is achieved by linear SVM. For password dataset, the minimum error rate raises to 11.6\% and corresponds to Linear discriminant analysis. The minimum error rate for the concatenated dataset is same as the value obtained with the Turkish dataset but with a different algorithm. This result shows that rather than applying a single algorithm to all contexts, it is better to select a suitable algorithm considering the dataset properties.

\section{PROTECTION AGAINST IMITATION}

One serious problem of a Keystroke Dynamics based age group predictor could be caused by an individual consciously altering his/her behavior during typing in order to falsify the classification. This is less of an issue for commercial applications. Similarly, if users are not aware that such an application is in operation (\emph{e.g.}, when it is used for a police investigation), then imitation is less likely. This behavior also seems less likely for children if they have been already typing at their best. But, in some cases it may be possible for some adults to imitate the typing behavior of children consciously.

To investigate this issue, we analyze whether the algorithms we employed are resistant to imitative behavior of adults. For this purpose, we have asked 20 adults to type the same two phrases like a primary school student but we did not give any clue about the typing behavior of children. The adults responded to our request by typing slower in general. Some of them used just one or two fingers for typing, some of them made random pauses during typing. Some others preferred to use CAPS LOCK key instead of SHIFT key for typing capital letters. We name the group of samples that are collected from these adults who try to imitate children's typing behavior as  ``impostor dataset''.

The impostor dataset is divided into 5 subsets, one of them is reserved as the test set and the remaining 4 subsets are used for training the classifier together with the training data used in the previous section\footnote{Inclusion of impostor data to the training is our implementation choice. Otherwise, error rates for non-impostors would stay as same at a cost of increased error rates for impostors.}. The error rates are listed in Table \ref{impostors}. The column with the heading ``Imp. Err.'' denotes the error rate for the impostors, which corresponds to the ratio of impostor test samples misclassified as a child sample to the total number of impostor test samples. When impostor samples are included in the training set the behavior of the classifier is also affected, leading to a change in equal error rates (EER) for non-impostors. Therefore we also present these new results in Table \ref{impostors}.

\begin{table*}[!htb]
\caption{Error rates when impostor samples are included.}
\label{impostors}
\centering
\renewcommand{\arraystretch}{1.2}
\begin{tabular}{|p{4.2cm}|>{\centering\arraybackslash}m{0.5in}|>{\centering\arraybackslash}m{0.5in}|>{\centering\arraybackslash}m{0.5in}|>{\centering\arraybackslash}m{0.5in}|>{\centering\arraybackslash}m{0.5in}|>{\centering\arraybackslash}m{0.5in}|}
\hline
 &  \multicolumn{2}{c|}{\textbf{Turkish dataset (\%) }} &  \multicolumn{2}{c|}{\textbf{Password dataset (\%) }} &  \multicolumn{2}{c|}{\textbf{Concatenated dataset (\%) }}\\
\hline
  \textbf{Algorithm} & \textbf{EER}   & \textbf{Imp. Err.}  & \textbf{EER}   & \textbf{Imp. Err.}  & \textbf{EER}   & \textbf{Imp. Err.}
  \\
\hline
Speed (Total time)	&	27.5	&	49.0	&	29.3	&	37.0	&	26.8	&	43.0	\\	\hline
Euclidean distance	&	21.2	&	53.0	&	21.2	&	33.0	&	20.0	&	41.0	\\	\hline
Manhattan distance	&	23.7	&	51.0	&	24.0	&	\textbf{32.0}	&	22.5	&	42.0	\\	\hline
Nearest neighbor	&	20.0	&	56.0	&	19.3	&	35.0	&	18.5	&	47.0	\\	\hline
Linear discriminant analysis	&	19.3	&	56.0	&	\textbf{18.0}	&	41.0	&	15.9	&	39.0	\\	\hline
Support vector machine (Linear) 	&	17.7	&	43.0	&	19.3	&	41.0	&	\textbf{15.7}	&	\textbf{28.0}	\\	\hline
Support vector machine (RBF)	&	\textbf{17.3}	&	\textbf{38.0}	&	19.9	&	37.0	&	18.8	&	41.0	\\	\hline
Gradient descent bp. 	&	47.7	&	67.0	&	47.5	&	53.0	&	59.5	&	56.0	\\	\hline
Conjugate gr. bp. with FR up.	&	23.5	&	63.0	&	21.7	&	36.0	&	22.3	&	54.0	\\	\hline
BFGS quasi-Newton bp.	&	22.7	&	59.0	&	23.5	&	40.0	&	24.2	&	54.0	\\	\hline
One step secant bp.	&	21.7	&	50.0	&	22.2	&	40.0	&	23.2	&	52.0	\\	\hline
Scaled conjugate gradient bp.	&	24.5	&	64.0	&	23.3	&	43.0	&	23.3	&	55.0	\\	\hline
Levenberg-Marquardt bp.	&	26.0	&	68.0	&	23.4	&	44.0	&	27.2	&	63.0	\\	\hline

\end{tabular}
\end{table*}

The error rates for the impostor samples are far from being promising. The minimum error rate for the impostor group is 28.0\%, which is the result of applying linear SVM to concatenated dataset. Moreover, the equal error rates for the genuine samples take the values between 15.7 to 18.0 percent when impostor samples are included in the training set. With these results, it can be concluded that Keystroke Dynamics based age group detection is vulnerable to non-zero effort attacks.

\section{FURTHER DISCUSSION}
We present here further discussion about the experimental results. The error rates presented in this study for discriminating age groups may be unacceptable for some of the forensic applications. However, we still believe that keystoke dynamics based age group detection can help in many investigations. Consider a scenario in which a criminal introduces himself as a child on an instant messaging application. A policeman who is suspicious of this malicious activity also logins to the application and introduces himself as a child to the suspect. In this scenario, if the police investigator is able to obtain the keystroke data from the other party, he can use a Keystroke Dynamics evaluator to make a prediction about the age group of the suspect. In order to improve his prediction accuracy, the investigator can also benefit from other information (\emph{e.g}, the words and sentences the other party uses). By combining several clues, s/he can make a valid guess about the age group of the suspect.

Unlike forensic applications, commercial applications are less sensitive to classification errors. We think the error rates obtained in this study are satisfactory for a commercial web site which displays products according to the age group of the user.

In addition to accuracy, an important aspect for the use of biometric information is privacy. Collecting personal information may cause frustration among users. Moreover, such an act may trigger legal disputes. With careful application choices, problems related to privacy can be mitigated for Keystroke Dynamics based age group detection. First, training data can be stored in de-identified format without personal information except age. Furthermore, if the training data will not be updated in the future, it has no use hence can be truncated after the classifiers are trained (except for the nearest neighbor classifier). In the test phase, the keystroke data of the user can be processed on the fly and there is no need for the storage of the data.

An issue yet to be discussed is the computation overhead of the algorithms that are used. Despite the negligible run times of distance metric algorithms, support vector machines and neural networks require considerable computation times for learning (in the order of milliseconds per user when run on a laptop PC). However, since the training process can be executed offline on the server side, it does not slow down the application. Classification time, which is the main factor that actually determines the run time of the application, is negligible for all the methods including neural networks and support vector machines.

A question that may be relevant for using the same phrase for all users is that whether it is realistic to expect users type the same phrase in a real life application. We think that fixed text requirement could be eliminated if a keystroke library is built by collecting a larger set of typing data as the test samples. In the application phase, incoming typing data can be compared against the samples in the library and a proper subset could be selected.

\section{CONCLUSION AND FUTURE WORK}

Using biometric characteristics for identifying individuals is becoming increasingly pervasive. Biometric data can also be used to determine common group characteristics. In this study, we show that Keystroke Dynamics, which refers to typing style of computer users, can be successfully used to predict whether the age of the computer user is below 15. Experimental results show that accuracy rates above 90 percent are achievable with a careful selection of classification methodology. On the other hand, we observe a significant increase in the error rates when there are human impostors in the system. We expect the error rates would increase even further with algorithmic forgeries \cite{rahman2013snoop}.


Other than the experimental findings, we believe that another significant contribution of this study is the collected dataset. Although there are a number of datasets that can be used for Keystroke Dynamics based user verification, our dataset is the first one that contains age information and made publicly available \cite{dataset}. Moreover, we also publish our data collection application and test scripts so that the study can be easily replicated by other researchers.

The results presented in this paper pave the way for future studies. A rapidly increasing number of consumers (especially youngsters) prefer smart phones with touch screens to access online applications, especially social networks. This imposes the interesting question whether Keystroke Dynamics can be used for age group detection when these devices are used. Hence, collecting a keystroke dataset on a handset application and classifying the typing patterns with respect to age groups will be a valuable study.

Similar to Keystroke Dynamics, Mouse Dynamics is an emerging field to authenticate computer users, which is based on timing, movement direction and clicking information during mouse use \cite{mouse}. At the moment, it remains as an open problem whether mouse data can be used to infer the age group of individuals. If it is shown that Mouse Dynamics can be used for the same purpose, an application that uses keystroke and mouse data together may provide more accurate results about the age group.

In addition to age group, it is of question whether it is possible to classify users based on some other characteristics such as nationality, left-handedness and even the height of individuals.

The age group information may increase the success rate of keystroke based authentication. It was shown that the accuracy of user authentication could be increased by 20 percent with gender prediction \cite{giot}. Similarly, an improvement can be achieved using age group prediction with keystroke data. A future work that explores this possibility may be another valuable contribution.

%

\bibliographystyle{IEEEtran}
\bibliography{IEEEabrv,InternetChildren_double}

\begin{thebibliography}{10}
\providecommand{\url}[1]{#1}
\csname url@samestyle\endcsname
\providecommand{\newblock}{\relax}
\providecommand{\bibinfo}[2]{#2}
\providecommand{\BIBentrySTDinterwordspacing}{\spaceskip=0pt\relax}
\providecommand{\BIBentryALTinterwordstretchfactor}{4}
\providecommand{\BIBentryALTinterwordspacing}{\spaceskip=\fontdimen2\font plus
\BIBentryALTinterwordstretchfactor\fontdimen3\font minus
  \fontdimen4\font\relax}
\providecommand{\BIBforeignlanguage}[2]{{%
\expandafter\ifx\csname l@#1\endcsname\relax
\typeout{** WARNING: IEEEtran.bst: No hyphenation pattern has been}%
\typeout{** loaded for the language `#1'. Using the pattern for}%
\typeout{** the default language instead.}%
\else
\language=\csname l@#1\endcsname
\fi
#2}}
\providecommand{\BIBdecl}{\relax}
\BIBdecl

\bibitem{india}
\BIBentryALTinterwordspacing
M.~Variyar. (2013, February) {82\% children on Facebook receive vulgar
  messages}. [Online]. Available:
  \url{http://www.hindustantimes.com/India-news/Mumbai/82-children-on-Facebook%
-receive-vulgar-messages/Article1-1017029.aspx}
\BIBentrySTDinterwordspacing

\bibitem{dataset}
\BIBentryALTinterwordspacing
(2014, October) Detecting age groups using keystroke dynamics. [Online].
  Available: \url{http://bil.etu.edu.tr/bicakci/dagkd/dagkd.htm}
\BIBentrySTDinterwordspacing

\bibitem{anomaly}
K.~S. Killourhy and R.~A. Maxion, ``Comparing anomaly-detection algorithms for
  keystroke dynamics,'' in \emph{Proceedings of the 39th Annual Dependable
  Systems and Networks Conference}.\hskip 1em plus 0.5em minus 0.4em\relax
  IEEE, 2009, pp. 125--134.

\bibitem{rlang}
\BIBentryALTinterwordspacing
{R Development Core Team}, \emph{R: A Language and Environment for Statistical
  Computing}, R Foundation for Statistical Computing, Vienna, Austria, 2008,
  {ISBN} 3-900051-07-0. [Online]. Available: \url{http://www.R-project.org}
\BIBentrySTDinterwordspacing

\bibitem{second}
Y.~Uzun and K.~Bicakci, ``A second look at the performance of neural networks
  for keystroke dynamics using a publicly available dataset.'' \emph{Computers
  and Security}, vol.~31, no.~5, pp. 717--726, 2012.

\bibitem{snoop}
K.~Rahman, K.~Balagani, and V.~Phoha, ``Making impostor pass rates meaningless:
  A case of snoop-forge-replay attack on continuous cyber-behavioral
  verification with keystrokes,'' in \emph{Computer Vision and Pattern
  Recognition Workshops (CVPRW), 2011 IEEE Computer Society Conference on},
  2011, pp. 31 --38.

\bibitem{rahman2013snoop}
K.~A. Rahman, K.~S. Balagani, and V.~V. Phoha, ``Snoop-forge-replay attacks on
  continuous verification with keystrokes,'' \emph{Information Forensics and
  Security, IEEE Transactions on}, vol.~8, no.~3, pp. 528--541, 2013.

\bibitem{traits}
A.~Serwadda, V.~V. Phoha, and A.~Kiremire, ``Using global knowledge of users'
  typing traits to attack keystroke biometrics templates,'' in
  \emph{Proceedings of the thirteenth ACM multimedia workshop on Multimedia and
  security}, ser. MMSec '11.\hskip 1em plus 0.5em minus 0.4em\relax New York,
  NY, USA: ACM, 2011, pp. 51--60.

\bibitem{application}
{Thornton, M. A.}, ``Keystroke dynamics,'' in \emph{Encyclopedia of
  Cryptography and Security}.\hskip 1em plus 0.5em minus 0.4em\relax Springer
  Publishers, 2011.

\bibitem{giot}
R.~Giot and C.~Rosenberger, ``A new soft biometric approach for keystroke
  dynamics based on gender recognition.'' \emph{International Journal of
  Information Technology and Management (IJITM)}, vol.~11, no. 1/2, pp. 35--49,
  2012.

\bibitem{graphology}
\BIBentryALTinterwordspacing
{Encyclopedia Britannica}. (2014, July 13) Graphology. [Online]. Available:
  \url{http://www.britannica.com/EBchecked/topic/242077/graphology}
\BIBentrySTDinterwordspacing

\bibitem{fingerprint}
``Estimation of age through fingerprints using wavelet transform and singular
  value decomposition,'' \emph{International Journal of Biometrics and
  Bioinformatics (IJBB)}, vol.~6, no.~2, pp. 58--67, 2012.

\bibitem{keyboard}
R.~Giot, M.~EI-Abed, and C.~Rosenberger, ``Keystroke dynamics with low
  constraints svm based passphrase enrollment,'' in \emph{Proceedings of the
  3rd IEEE international conference on Biometrics: Theory, applications and
  systems}, ser. BTAS'09.\hskip 1em plus 0.5em minus 0.4em\relax Piscataway,
  NJ, USA: IEEE Press, 2009, pp. 425--430.

\bibitem{free}
K.~S. Killourhy and R.~A. Maxion, ``Free vs. transcribed text for
  keystroke-dynamics evaluations,'' in \emph{Proceedings of the 2012 Workshop
  on Learning from Authoritative Security Experiment Results}, ser. LASER
  '12.\hskip 1em plus 0.5em minus 0.4em\relax New York, NY, USA: ACM, 2012, pp.
  1--8.

\bibitem{matlab}
MATLAB, \emph{version 7.6.0}.\hskip 1em plus 0.5em minus 0.4em\relax Natick,
  Massachusetts: The MathWorks Inc., 2010.

\bibitem{duda}
R.~O. Duda, P.~E. P.~E. Hart, and D.~G. Stork, \emph{Pattern Classification},
  2nd~ed.\hskip 1em plus 0.5em minus 0.4em\relax Wiley, 2001.

\bibitem{tutorial}
C.~J.~C. Burges, ``A tutorial on support vector machines for pattern
  recognition,'' \emph{Data Mining and Knowledge Discovery}, vol.~2, no.~2, pp.
  121--167, 1998.

\bibitem{toolbox}
H.~Demuth and M.~Beale, \emph{Neural Network Toolbox 7 User’s Guide}, The
  MathWorks, Inc, 2010.

\bibitem{mouse}
K.~Revett, H.~Jahankhani, S.~T. Magalh\~{a}es, and H.~M.~D. Santos, \emph{{A
  Survey of User Authentication Based on Mouse Dynamics}}, ser. Communications
  in Computer and Information Science.\hskip 1em plus 0.5em minus 0.4em\relax
  Berlin, Heidelberg: Springer Berlin Heidelberg, 2008, vol.~12, ch.~25, pp.
  210--219.

\end{thebibliography}

\end{document}